\documentclass[prb,showpacs,superscriptaddress,twocolumn]{revtex4}

\usepackage{mathrsfs}
\usepackage{graphicx}
\usepackage{amsfonts}
\usepackage{amsmath}
\usepackage{amssymb}

\newcommand{\gv}[1]{\ensuremath{\mbox{\boldmath$ #1 $}}} 
\newcommand{\abs}[1]{\left| #1 \right|} 
\newcommand{\grad}[1]{\gv{\nabla} #1} 
\newcommand{\vint}[1]{\int\mathrm{d}^3 #1\,}

\def\up{\uparrow}
\def\down{\downarrow}

\newcommand{\normG}[2]{\ensuremath{\mathscr{G}_{\nu,\omega_n}^{(0)}(\mathbf{#1},\mathbf{#2})}}
\newcommand{\tildeG}[2]{\ensuremath{\widetilde{\mathscr{G}}_{\nu,\omega_n}^{(0)}(\mathbf{#1},\mathbf{#2})}}
\newcommand{\G}[2]{\mathscr{G}_{\nu,\omega_n}(\mathbf{#1},\mathbf{#2})}
\newcommand{\F}[2]{\mathscr{F}_{\nu,\omega_n}^\dag(\mathbf{#1},\mathbf{#2})}

\newcommand{\D}[1]{\Delta_\nu (\mathbf{#1})}
\newcommand{\Ds}[1]{\Delta_\nu^* (\mathbf{#1})}

\begin{document}

\title[]{Extended Uniform Ginzburg-Landau Theory for Novel Multiband Superconductors}
\today

\author{Brendan J. Wilson}
\email{brendan.wilson@anu.edu.au}

\author{Mukunda P. Das}
\email{mukunda.das@anu.edu.au}

\affiliation{Department of Theoretical Physics, Research School of Physics and Engineering,
The Australian National University, Canberra ACT 0200, Australia}

\begin{abstract}
The recently discovered multiband superconductors have created a new class of novel superconductors. In these materials multiple superconducting gaps arise due to the formation of Cooper pairs on different sheets of the Fermi surfaces.  An important feature of these superconductors is the interband couplings, which not only change the individual gap properties, but also create new collective modes.
Here we investigate the effect of the interband couplings in the Ginzburg-Landau theory. We produce a general $\tau^{(2n+1)/2}$ expansion  ($\tau = 1-T/T_c$) and  show that this expansion has unexpected behaviour for $n\geq 2$. This point emphasises the weaker validity of the GL theory for lower temperatures and gives credence to the existence of hidden criticality near the critical temperature of the uncoupled subdominant band.
\end{abstract}

\pacs{74.20.De, 74.20.Fg, 74.50.+r}

\maketitle

\section{Introduction}

The BCS theory is considered as a conventional theory of superconductivity at a microscopic level, where bosonic excitations like phonons or spin fluctuations play the role of mediators in the formation of Cooper pairs and hence superconductivity occurs in a metallic state. In recent years many varieties of superconductors have been discovered which are unconventional. Some examples are 
\begin{enumerate}
\item cuprates, where anisotropic d-wave gaps occur with nodes (vanishing gap) at some symmetry points on the Fermi surface \cite{Yoshida,Yoshida2}, 
\item co-existence of superconductivity with anti-ferromagnetism (CeCuSi$_2$) or with ferromagnetism (UGe$_2$) predominantly in heavy fermion systems \cite{Knebel, Bert}, 
\item coexistence of superconductivity with charge/spin ordering (NbSe$_2$, Cuprates) \cite{Gabovich}, 
\item non-centrosymmetric heavy fermion systems (CePt$_3$Si, CeInSi$_3$) \cite{Kimura}, where lack of inversion symmetry gives rise to spin-orbit interaction with no definite parity in the ground state. In this case singlet and triplet pairings coexist,
\item strong electronic correlations dominated so-called non-fermi liquid state, believed to be found in high $T_c$ oxides and in heavy fermion systems \cite{Stewart}.
\end{enumerate}
In addition to these, there is another class of unconventional superconducting systems, the novel multiband superconductors. In these systems two or more energy bands are cut by the fermi energy giving rise to multiple energy gaps with different magnitudes in the different Fermi sheets. Recent measurements of tunnelling, point-contact spectroscopy, angle-resolved photoemission and specific heat provide clear evidence of multiple gap structures. Examples of such systems are MgB$_2$, RNi$_2$B$_2$C (R= Lu,Y), 2H-NbSe$_2$ and many in the pnictide FeAs family.

The BCS theory of superconductivity has been generalised to multiband systems. In one of our recent publications we have given a brief appraisal of the history of multiband BCS theory \cite{Wilson} and have presented a theory of the time-reversal symmetry broken state in the BCS formalism. We are reminded that a phenomenological theory of superconductivity by Ginzburg and Landau [GL] was developed before the proposal of microscopic BCS theory. The GL approach is a successful theory of phase transitions with many practical applications. The basis of this theory rests on two approximations correct around the critical temperature ($T_c$) (i) the order parameter, $\Psi$, is small near $T_c$  and (ii) $\Psi\sim  \tau^{1/2}$, where $\tau=1- (T/T_c)$.  Despite these limitations there is a myth that the GL theory applies not only around $T_c$, rather it is useful for much lower temperatures.  In early years soon after the appearance of BCS theory, Gor'kov \cite{Gorkov, Abrikosov} established the equivalence of the BCS energy gap, $\Delta$, with the order parameter of the GL theory for single band superconductors with the above conditions (i) and (ii).  

In multiband superconductors the equivalence has been investigated by many authors (see a brief review in ref. \cite{Brandt}). In both theories (BCS and GL) an additional interaction term appears due to interband interaction, which is recognised as the Josephson term.  This term is the lowest order coupling between the gaps (in BCS) and order parameter (in GL) in the different bands.  The presence of Josephson terms in multiband superconductors causes several problems in the Gor'kov type derivations. Recently in a series of papers Vagov and coworkers \cite{Shanenko,Vagov1,Vagov2,Vagov3} have made detailed analysis and established a generalisation of the standard GL theory (which is correct to $\tau^{1/2}$) by retaining additional terms in the expansion up to order $\tau^{(2n+1)/2}$. In practice they have analysed the $n=1$ corrections to the order parameter for 1, 2 and 3 bands . They call this formalism extended GL theory. This extended version with $\tau^{3/2}$ corrections seems to have improved the validity of the GL expansion to some lower temperatures away from $T_c$ in one- and two-band systems.

In this paper we adopt the microscopic approach of Gor'kov generally for uniform multiband systems with isotropic (spherical) Fermi surfaces. Other types of  Fermi surfaces for dirty superconductors and with anisotropy can be done appropriately with more complications. We present here our detailed calculations of BCS gaps and GL order parameters for superconductors with one and two bands.  

In Sec.2 we extend the Gor'kov technique to multiband superconductors in the absence of an external magnetic field, going beyond the standard/traditional model of GL. The coefficients for all terms in the series expansion of the self-consistent gap equation are given explicitly, and we show how to solve the resulting equations for the gap functions.

In Sec.3 results for the one band superconductor are presented showing clearly the departure of the standard GL order parameter (with $\tau^{1/2}$) while comparing with the BCS result.  Higher order corrections are reported with impressive agreement with the BCS. We see that each additional term increases the range of $\tau$ for which the expansion is accurate.

In Sec.4 similar results for the two band superconductors are presented. These calculations are done for different interband couplings. In contrast with the single band results, additional terms in the two-band GL expansion only improve the agreement with the BCS result up to a certain value for $\tau$. Pushing beyond this point, the agreement becomes worse as additional terms are added. This disagreement is associated with the appearance of a second critical temperature in the weak coupling limit.

In Sec.5  we present the conclusions and summary of this work.

\section{Derivation of Extended Ginzburg-Landau Theory}
The BCS theory is generalised to a multiband theory \cite{Suhl,Moskalenko} by allowing multiple fermion operators, which are identified by a band index, $\nu$, and including a Josephson interband term in the interaction. This term allows for Cooper pairs to tunnel from band to band. With this generalisation, the effective multiband BCS Hamiltonian in real space is given by
\begin{align}
&\hat{H}_{eff}=\sum_\sigma\sum_\nu\vint{x} \hat{\psi}^\dag_{\sigma \nu}(\mathbf{x})\left(\Pi_\nu(\mathbf{x})-\mu_\nu\right)\hat{\psi}_{\sigma \nu}(\mathbf{x})\nonumber\\
&+\sum_{\nu}\vint{x} \left(\Delta_\nu^*(\mathbf{x})\hat{\psi}_{\down \nu}(\mathbf{x})\hat{\psi}_{\up \nu}(\mathbf{x})+\mathrm{h.c.}\right),
\end{align}
where $\Pi_\nu(\mathbf{x})=\frac{1}{2m_\nu}\left(-i\hbar\grad-\frac{e\mathbf{A}(\mathbf{x})}{c}\right)^2$, $\mathbf{A}(\mathbf{x})$ is the vector potential, $\hat{\psi}_{\sigma \nu}$ ($\hat{\psi}^\dag_{\sigma \nu}$) are fermionic annihilation (creation) operators, $\nu$, $\nu'$ are band indices, $\sigma$ are spin indices, $m_\nu$ is the electron mass, $\mu_\nu$ is the chemical potential, $g_{\nu\nu'}$ are the interband coupling parameters, and $\Delta_{\nu}(\mathbf{x})=\sum_{\nu'}g_{\nu\nu'}\left<\hat{\psi}_{\nu'\up}(\mathbf{x})\hat{\psi}_{\nu'\down}(\mathbf{x})\right>$ is the superconducting gap.

Following the Gor'kov technique, the Green function $\G{x}{x'}$  and anomalous Green function $\F{x}{x'}$ can be written as a pair of coupled integral equations \cite{Gorkov,Abrikosov}:
\begin{align}
\G{x}{x'}=&\normG{x}{x'}\nonumber\\
&-\hbar^{-1}\vint{y}\normG{x}{y}\D{y}\F{y}{x'}\label{eq:NormalGreenFunction}\\
\F{x}{x'}=&\hbar^{-1}\vint{y}\tildeG{x}{y}\Ds{y}\G{y}{x'},\label{eq:AnomalousGreenFunction}
\end{align}
where $\normG{x}{y}$ is the normal Green function and $\tildeG{x}{y}={\mathscr{G}_{\nu,-\omega_n}^{(0)}(\mathbf{y},\mathbf{x})}$, $\D{x}$ is the superconducting gap function in band $\nu$, and the fermionic Matsubara frequency $\omega_n=(2n+1)\frac{\pi}{\beta\hbar}$, with $\beta=1/k_B T$. The normal Green functions satisfy the equations
\begin{align}
\left[i\hbar\omega_n+\frac{\hbar^2}{2m_\nu}\left(\nabla+\frac{ie\mathbf{A}(\mathbf{x})}{\hbar c}\right)^2+\mu_\nu\right]&\normG{\mathbf{x}}{\mathbf{x'}}\nonumber\\
&=\hbar\delta^3(\mathbf{x}-\mathbf{x'})\\
\left[-i\hbar\omega_n+\frac{\hbar^2}{2m_\nu}\left(\nabla-\frac{ie\mathbf{A}(\mathbf{x})}{\hbar c}\right)^2+\mu_\nu\right]&\tildeG{\mathbf{x}}{\mathbf{x'}}\nonumber\\
&=\hbar\delta^3(\mathbf{x}-\mathbf{x'}).
\end{align}
Using substitution, we can transform equations (\ref{eq:NormalGreenFunction}) and (\ref{eq:AnomalousGreenFunction}) into decoupled nonlinear integral equations, and by continued substitution we can write the anomalous Green function as a series expansion in the gap and the normal Green function
\begin{align}
&\F{x}{x'}=\nonumber\\
&\sum_{m=0}^{\infty}\frac{(-1)^{m}}{\hbar^{2m+1}}\left(\prod_{j=1}^{2m+1}\vint{y_{i}}\right)\tildeG{x}{y_1}\Ds{y_{1}}\nonumber\\
&\times\left(\prod_{j=1}^{m}\normG{y_{2j-1}}{y_{2j}}\D{y_{2j}}\tildeG{y_{2j}}{y_{2j+1}}\Ds{y_{2j+1}}\right)\nonumber\\
&\times\normG{y_{2m+1}}{x'}.
\end{align}

The gap is defined in terms of the anomalous Green function by
\begin{align}
\sum_{\nu'} \left[g^{-1}\right]_{\nu\nu'}\Delta^*_{\nu'}(\mathbf{x})=&\lim_{\eta\to 0^+}\sum_n e^{-i\omega_n\eta}\frac{1}{\beta\hbar}\F{x}{x}.
\label{eq:gapdefinition}
\end{align}

\subsection{Uniform field free case}
In this paper we are interested in finding the mean value for the gaps, so we will consider the case where the magnetic field is zero and the gap does not depend on $\mathbf{x}$, so the superconductor is uniform.
By requiring the gap to satisfy equation \ref{eq:gapdefinition}, we obtain the self-consistent gap equation in matrix form
\begin{equation}
\check{g}^{-1}.\vec{\Delta}=\vec{R},
\label{eq:selfconsistentgap}
\end{equation}
where $\check{g}$ is the interband coupling matrix with elements $g_{\nu,\nu'}$, $\vec{\Delta}$ is a column vector with elements $\Delta_\nu$, and $\vec{R}$ is a column vector with elements given by
\begin{align}
R_\nu=&\sum_{m=0}^{\infty}\Delta_\nu\abs{\Delta_\nu}^{2m}P_{\nu,m},\\
P_{\nu,m}=&\lim_{\mathbf{y_{2m+2}}\to\mathbf{y_0}}\Bigg(\prod_{j=1}^{2m+1}\vint{y_j}\Bigg)Q_{\nu,m}(\{\mathbf{y}\}_{2m+2}),
\label{eq:P}
\end{align}
\begin{align}
&Q_{\nu,m}(\{\mathbf{y}\}_{2m+2})=\nonumber\\
&\frac{(-1)^m}{\beta\hbar^{2(m+1)}}\sum_n \prod_{j=1}^{m+1}\normG{y_{2j-2}}{y_{2j-1}}\tildeG{y_{2j-1}}{y_{2j}},
\end{align}
with $\{\mathbf{y}\}_{m}=\{\mathbf{y_0},\mathbf{y_1}\ldots,\mathbf{y_m}\}$.
Equation \ref{eq:selfconsistentgap} is a coupled equation involving the gaps from all bands, $\nu$. These equations must be solved simultaneously.

The normal Green function can be solved in Fourier space to find
\begin{align}
\normG{y}{x}=&\frac{1}{(2\pi)^3}\vint{k}\frac{\hbar e^{i\mathbf{k}.(\mathbf{y}-\mathbf{x})}}{i\hbar\omega_n-\xi_{\nu,k}},
\end{align}
with $\xi_{\nu,k}=\frac{\hbar^2k^2}{2m_\nu}-\mu_\nu$. 
Performing each of the real space integrals in equation \ref{eq:P} produces a delta function, and these can be used to compute all but one of the $k$ space integrals, resulting in the simplified expression
\begin{align}
P_{\nu,m}=& 
\frac{(-1)^m}{\beta}\sum_{n=-\infty}^{\infty} \vint{k}\frac{1}{((\hbar\omega_n)^2+\xi_{\nu,k}^2)^{m+1}}\nonumber\\
=&\frac{(-1)^m}{\beta}N_\nu(0)\sum_{n=-\infty}^{\infty}\int\mathrm{d}\xi\,\frac{1}{((\hbar\omega_n)^2+\xi^2)^{m+1}},
\end{align}
with  $N_\nu (0)$ is the density of states in band $\nu$.
When $m=0$ this integral diverges logarithmically, and so must be cut off at the Debye energy, $\hbar\omega_D$. In this case we find
\begin{align}
P_{\nu,0}
=&\beta^{-1}N_\nu(0)\sum_{n=-\infty}^{\infty}\int_{-\hbar\omega_D}^{\hbar\omega_D}\mathrm{d}\xi\,\frac{1}{(\hbar\omega_n)^2+\xi^2}\nonumber\\
\approx&N_\nu(0)\mathcal{A}-a_\nu\ln\left(\frac{1}{1-\tau}\right),\\
\mathcal{A}=&\ln\left(\frac{2\hbar\omega_D e^\gamma}{\pi k_B T_c}\right),\\
a_\nu=&-N_\nu(0),
\end{align}
where $\gamma\approx0.577216$ is the Euler-Mascheroni constant and $\tau=1-T/T_c$ with $T_c$ to be defined later.
The remaining terms with $m\geq1$ may be computed directly
\begin{align}
P_{\nu,m}=& 
\frac{(-1)^m}{\beta}N_\nu(0)\sum_{n=-\infty}^{\infty}\int_{-\infty}^{\infty}\mathrm{d}\xi\,\frac{1}{((\hbar\omega_n)^2+\xi^2)^{m+1}}\nonumber\\
=&-b_{\nu,m}\frac{1}{(1-\tau)^{2m}},\\
b_{\nu,m}=&-N_\nu(0)\frac{(-1)^m \left(2^{2m+1}-1\right)(2m)!\zeta(2m+1)}{(4\pi)^{2m}(m!)^2(k_B T_c)^{2m}} ,
\end{align}
where  $\zeta(z)$ is the Riemann zeta function. Putting this back together we find
\begin{align}
R_\nu=& N_\nu(0)\mathcal{A}\Delta_\nu-a_\nu\ln\left(\frac{1}{1-\tau}\right)\Delta_\nu\nonumber\\
&-\sum_{m=1}^{\infty}b_{\nu,m} \frac{1}{(1-\tau)^{2m}}\abs{\Delta_\nu}^{2m}\Delta_\nu.
\end{align}

We then regroup terms to rewrite equation \ref{eq:selfconsistentgap} in the form
\begin{align}
0=&\check{L}.\vec{\Delta} +\vec{W},\label{eq:NewSelfConsistant}\\
W_\nu=&a_\nu\ln\left(\frac{1}{1-\tau}\right)\Delta_\nu+\sum_{m=1}^{\infty}b_{\nu,m} \frac{1}{(1-\tau)^{2m}}\abs{\Delta_\nu}^{2m}\Delta_\nu,
\end{align}
with $\check{L}=\check{g}^{-1}-\check{N}(0)\mathcal{A}$, and $\check{N}(0)$ is a diagonal matrix with elements $N_\nu (0)$ on the diagonal.

\subsection{Expansion in small $\tau$}
Near the transition temperature, $\tau$ is a small parameter, so we will expand equation \ref{eq:NewSelfConsistant} in powers of $\tau$. To truncate this expansion, keeping only terms up to $O\left(\tau^{(2n+1)/2}\right)$, we first make the scaling 
\begin{align}
\Delta_\nu=\tau^{1/2}\bar{\Delta}_\nu.
\end{align}
After scaling and then dividing through by $\tau^{1/2}$ we find
\begin{align}
0=&\check{L}.\vec{\bar{\Delta}} +\vec{\bar{W}},\label{eq:LWGap}\\
\bar{W}_\nu=&a_\nu\ln\left(\frac{1}{1-\tau}\right)\bar{\Delta}_\nu+\sum_{m=1}^{\infty}b_{\nu,m} \frac{\tau^m}{(1-\tau)^{2m}}\abs{\bar{\Delta}_\nu}^{2m}\bar{\Delta}_\nu.
\end{align}
Then the gap is expanded in powers of $\tau$, as is all the other dependence on $\tau$ in $\bar{W}_\nu$. The $\bar{\Delta}_{\nu}$ and $\bar{W}_{\nu}$ expansions are given by
\begin{align}
\bar{\Delta}_\nu(\tau)=\sum_{n=0}^{\infty}\bar{\Delta}_{\nu}^{(n)}\tau^n,\\
\bar{W}_\nu=\sum_{p=1}^{\infty}\bar{W}_{\nu}^{(p)}\tau^p.
\end{align}

We recover a set of equations for $\bar{\Delta}_\nu^{(n)}$ by collecting powers of $\tau$ in equation \ref{eq:LWGap} and requiring that the equality holds for all $\tau$.
The leading order behaviour  is a constant. Collecting these constant terms leads to the lowest order equation
\begin{align}
0=&\check{L}.\vec{\bar{\Delta}}^{(0)}.
\label{eq:TcEquation}
\end{align}
This has a non-trivial solution if $\det{\check{L}}=0$. We choose $T_c$ to be the largest value such that this equation is satisfied. We note that the $T_c$ of the combined system depends on the interband coupling. In the one band case, the well known solution for $T_c$ is 
\begin{align}
T_c=&\frac{2 e^\gamma\hbar\omega_D}{\pi}\exp\left(-\frac{1}{g_{11}N_1(0)}\right),
\end{align}
while for the two-band case, the solution for $T_c$ is 
\begin{widetext}
\begin{align}
T_c=&\frac{2 e^\gamma\hbar\omega_D}{\pi}\exp{\left(-\frac{g_{11}N_1(0)+g_{22}N_2(0)-\sqrt{(g_{11}N_1(0)-g_{22}N_2(0))^2+4g_{12}^2N_1(0)N_2(0)}}{2(g_{11}g_{22}-g_{12}^2)N_1(0)N_2(0)}\right)}\\
\approx&T_{1c}\left(1+\frac{g_{12}^2N_2(0)}{g_{11}^2N_1(0)(g_{11}N_1(0)-g_{22}N_2(0))}+O\left(g_{12}^4\right)\right),
\end{align}
\end{widetext}
when  $g_{11}N_1(0)>g_{22}N_2(0)$, where $T_{1c}$ is the critical temperature of the uncoupled first band, which is assumed to be the dominant band. We note that the critical temperature is enhanced over that of the dominant band due to the interband coupling, regardless of sign.

Now, since $\det{\check{L}}=0$, there is at least one eigenvector of $\check{L}$ with a zero eigenvalue. We shall assume that this is non-degenerate, so that there is only one zero eigenvalue. We choose the base eigenvector to have the form
\begin{align}
\vec{\eta}_1=&[\rho_1,\rho_2,\ldots,\rho_{N}]^T,\\
\rho_{i}=&\frac{c_{1,i}}{c_{1,1}},\\
c_{ijk\ldots,lmn\ldots}=&(-1)^{i+j+k+\ldots+l+m+n+\ldots}M_{ijk\ldots,lmn\ldots},
\end{align}
where $c_{ijk\ldots,lmn\ldots}$ is the cofactor of the matrix $\check{L}$, and $M_{ijk\ldots,lmn\ldots}$ is the minor of $\check{L}$, defined as the determinant of the matrix obtained by removing the rows $i,j,k,\ldots$ and columns $l,m,n,\ldots$ from $\check{L}$. Assuming all $\rho_i$ are finite and nonzero, we can then obtain a complete basis with the remaining vectors
\begin{align}
\vec{\eta}_i=[\rho_1,\rho_2,\ldots,-\rho_i,\ldots,\rho_{N}]^T.
\end{align}
The superconducting gaps can be written with this basis as
\begin{align}
\vec{\bar{\Delta}}^{(n)}=\sum_j \psi^{(n)}_{j}\vec{\eta}_j.
\label{eq:DeltaPsiBasis}
\end{align}
Putting this back into equation \ref{eq:TcEquation} and using the fact that $\check{L}.\vec{\eta}_1=0$ and $\check{L}.\vec{\eta}_j\neq 0$, $j\neq1$, we find
\begin{align}
\psi^{(0)}_{j}=0,\quad j\neq1,\\
\vec{\bar{\Delta}}^{(0)}=\psi^{(0)}_{1}\vec{\eta}_{1}.
\end{align}
where $\psi_1^{(0)}$ is yet to be determined.
The term linear in $\tau$ gives the equation
\begin{align}
0=&\check{L}.\vec{\bar{\Delta}}^{(1)}+\vec{\bar{W}}^{(1)},\label{eq:LinearInTau}\\
\bar{W}_\nu^{(1)}=& a_\nu\bar{\Delta}_\nu^{(0)}+b_{\nu,1}\bar{\Delta}_\nu^{(0)}\abs{\bar{\Delta}_\nu^{(0)}}^2.
\end{align}
This mixes $\vec{\bar{\Delta}}^{(0)}$ with $\vec{\bar{\Delta}}^{(1)}$, however, as pointed out in ref \onlinecite{Vagov2}, we can remove the $\vec{\bar{\Delta}}^{(1)}$ dependence using the fact that $\vec{\eta}_1^T.\check{L}=0$.
Projecting this equation on to $\vec{\eta}_1$ and using the solution for $\vec{\bar{\Delta}}^{(0)}$ we find
\begin{align}
0=&\sum_{\nu} a_\nu\eta_{1,\nu}^2\psi_1^{(0)}+b_{\nu,1}\eta_{1,\nu}^4\psi_1^{(0)}\abs{\psi_1^{(0)}}^2\\
=&a\psi_1^{(0)}+b_1\psi_1^{(0)}\abs{\psi_1^{(0)}}^2,
\end{align}
with $a=\sum_{\nu} a_\nu\eta_{1,\nu}^2$ and $b_1=\sum_{\nu}b_{\nu,1}\eta_{1,\nu}^4$. This has the same form as the one band uniform G-L equation.
Kogan and Schmalian \cite{Kogan} pointed out that the gradient term is also the same as the one band G-L equation, and thus there is only one coherence length near $T_c$, and the order parameters are proportional to each other.

Projecting equation \ref{eq:LinearInTau} onto the other basis vectors, $\vec{\eta}_i$, results in a further set of equations for the higher components, $\psi_j^{(1)}$. 
\begin{align}
0=&\left(\sum_{j\neq1} \vec{\eta}_{i}^{T}.\check{L}.\vec{\eta}_j\psi_j^{(1)}\right)\nonumber\\
&+\sum_{\nu} a_\nu\eta_{i,\nu}\eta_{1,\nu}\psi_1^{(0)}+b_{\nu,1}\eta_{i,\nu}\eta_{1,\nu}^3\psi_1^{(0)}\abs{\psi_1^{(0)}}^2\\
=&\sum_{j\neq1} \gamma_{ij}\psi_j^{(1)} +\alpha_i\psi_1^{(0)}+\beta_{i,1}\psi_1^{(0)}\abs{\psi_1^{(0)}}^2,
\end{align}
with $\gamma_{ij}=\vec{\eta}_{i}.\check{L}.\vec{\eta}_j$, $\alpha_i=\sum_{\nu} a_\nu\eta_{i,\nu}\eta_{1,\nu}=a-2a_{i}\rho_i^2$ and $\beta_{i,1}=\sum_{\nu} b_{\nu,1}\eta_{i,\nu}\eta_{1,\nu}^3=b-2b_{i,1}\rho_i^4$. The indices $i$ and $j$ refer to the basis vectors, $\vec{\eta}_j$, not the band indices, $\nu$.

This process can be continued recursively to find the G-L approximation to any order. We provide the form for the terms $\bar{W}_\nu^{(2)}$ and $\bar{W}_\nu^{(3)}$.
\begin{align}
\bar{W}_\nu^{(2)}=& a_\nu\Delta_\nu^{(1)}+b_{\nu,1}\left(2\Delta_\nu^{(1)}\abs{\Delta_\nu^{(0)}}^2+\Delta_\nu^{(0)2}\Delta_\nu^{(1)*}\right)\nonumber\\
&+ \frac{1}{2}a_\nu\Delta_\nu^{(0)}+2b_{\nu,1}\Delta_\nu^{(0)}\abs{\Delta_\nu^{(0)}}^2+b_{\nu,2}\Delta_\nu^{(0)}\abs{\Delta_\nu^{(0)}}^4,
\end{align}
\begin{align}
\bar{W}_\nu^{(3)}=& a_\nu\Delta_\nu^{(2)}+b_{\nu,1}\left(2\Delta_\nu^{(2)}\abs{\Delta_\nu^{(0)}}^2+\Delta_\nu^{(0)2}\Delta_\nu^{(2)*}\right)\nonumber\\
&+b_{\nu,1}\left(2\Delta_\nu^{(0)}\abs{\Delta_\nu^{(1)}}^2+\Delta_\nu^{(1)2}\Delta_\nu^{(0)*}\right)\nonumber\\
&+\frac{1}{2}a_\nu\Delta_\nu^{(1)}+2b_{\nu,1}\left(2\Delta_\nu^{(1)}\abs{\Delta_\nu^{(0)}}^2+\Delta_\nu^{(0)2}\Delta_\nu^{(1)*}\right)\nonumber\\
&+b_{\nu,2}\left(3\Delta_\nu^{(1)}\abs{\Delta_\nu^{(0)}}^4+2\abs{\Delta_\nu^{(0)}}^2\Delta_\nu^{(0)2}\Delta_\nu^{(1)*}\right)\nonumber\\
&+ \frac{1}{3}a_\nu\Delta_\nu^{(0)}+3b_{\nu,1}\Delta_\nu^{(0)}\abs{\Delta_\nu^{(0)}}^2+4b_{\nu,2}\Delta_\nu^{(0)}\abs{\Delta_\nu^{(0)}}^4\nonumber\\
&+b_{\nu,3}\Delta_\nu^{(0)}\abs{\Delta_\nu^{(0)}}^6.
\end{align}
All higher order terms can similarly be produced from the full definition of $\bar{W}_\nu$.

\section{Single-Band Ginzburg-Landau Theory}

\begin{figure}[t!]
\centering
\includegraphics[width=0.5\textwidth]{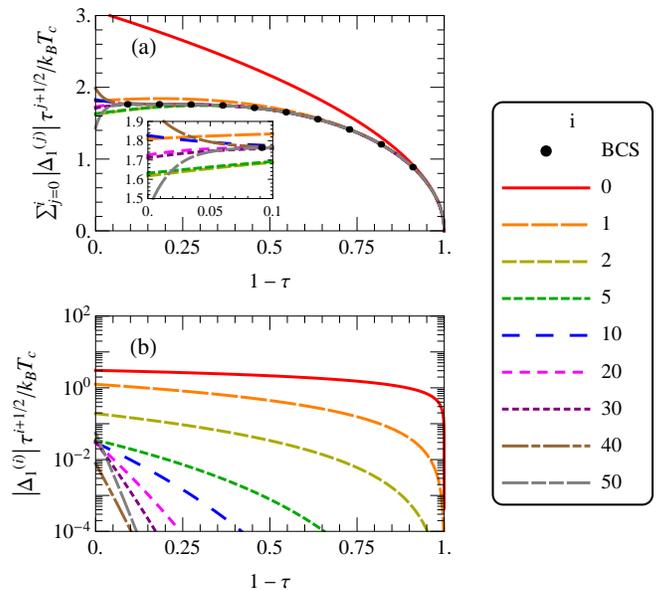}
\caption{(Colour online) (a) The extended GL expansion is compared to a numerical calculation of the full BCS result. The extended G-L converges to the true solution on the region $\tau<1$ and for moderate $\tau$ it converges quickly to the BCS solution. Inset: A close up of the region near $\tau=1$. There are singularities in the BCS function infinitesimally close to $\tau=1$ which prevent the extended G-L from converging at this point. (b) The magnitude of the lowest terms in the G-L expansion are shown on a Log plot. The magnitude of the higher terms decays quickly except near the point $\tau=1$ where it remains finite. This shows that the expansion is converging on the region $\tau<1$.}
\label{fig:ExtendedGLOneBandBigPlot}
\end{figure}

Applying this procedure to a single band superconductor is fairly straight forward. The matrix $\check{L}$ becomes a number, and the equation for $T_c$ becomes trivial to solve. The basis vector $\eta_1=1$ so that $\bar{\Delta}^{(n)}=\psi_1^{(n)}$ in equation \ref{eq:DeltaPsiBasis}. 

This procedure has been performed for the one band case to high order, with the results shown in figure \ref{fig:ExtendedGLOneBandBigPlot}.
The BCS solution is given by the bold black dots in the top plot. The thin red line that overshoots this is the conventional $\tau^{1/2}$ GL theory, while a selection of plots with higher order corrections up to $\tau^{(2n+1)/2}$ with $n=50$ are also shown. The first correction, $\tau^{3/2}$ is seen as the dashed line just above the BCS solution \cite{Vagov1}, while higher order corrections are almost indistinguishable except near $\tau=1$. Including a larger number of corrections increases the range of convergence, and it is presumed that the infinite sum will converge for all $\tau<1$. However for any large finite sum, the deviation near $\tau=1$ is expected to remain large.

On the bottom plot of figure \ref{fig:ExtendedGLOneBandBigPlot} we plot the magnitude of each term in the sum. The error of any finite sum is approximately given by the magnitude of the next term in the sum, and so this plot can be viewed as an estimation of the error in any given finite sum. The magnitude of each term decreases in general except near $\tau=1$, where, after the first few terms, it remains approximately constant.

For the single band case, an exact form for each term in the expansion can be computed, though the number of terms needed increases rapidly. We report the result for the first three terms in the expansion.
\begin{align}
\Delta^{(0)}_1=&k_BT_c\sqrt{\frac{8\pi^2}{7\zeta(3)}}\nonumber\\
\Delta^{(1)}_1=&\Delta^{(0)}_1\left(-\frac{3}{4}+\frac{93\zeta(5)}{196\zeta(3)^2}\right)\nonumber\\
\Delta^{(2)}_1=&\Delta^{(0)}_1\left(-\frac{11}{96}-\frac{93\zeta(5)}{784\zeta(3)^2}+\frac{8649\zeta(5)^2}{10976\zeta(3)^4}-\frac{635\zeta(7)}{1372\zeta(3)^3}\right).
\end{align}

\section{Two-Band Ginzburg-Landau Theory}

\begin{figure*}
\centering
\includegraphics[width=\textwidth]{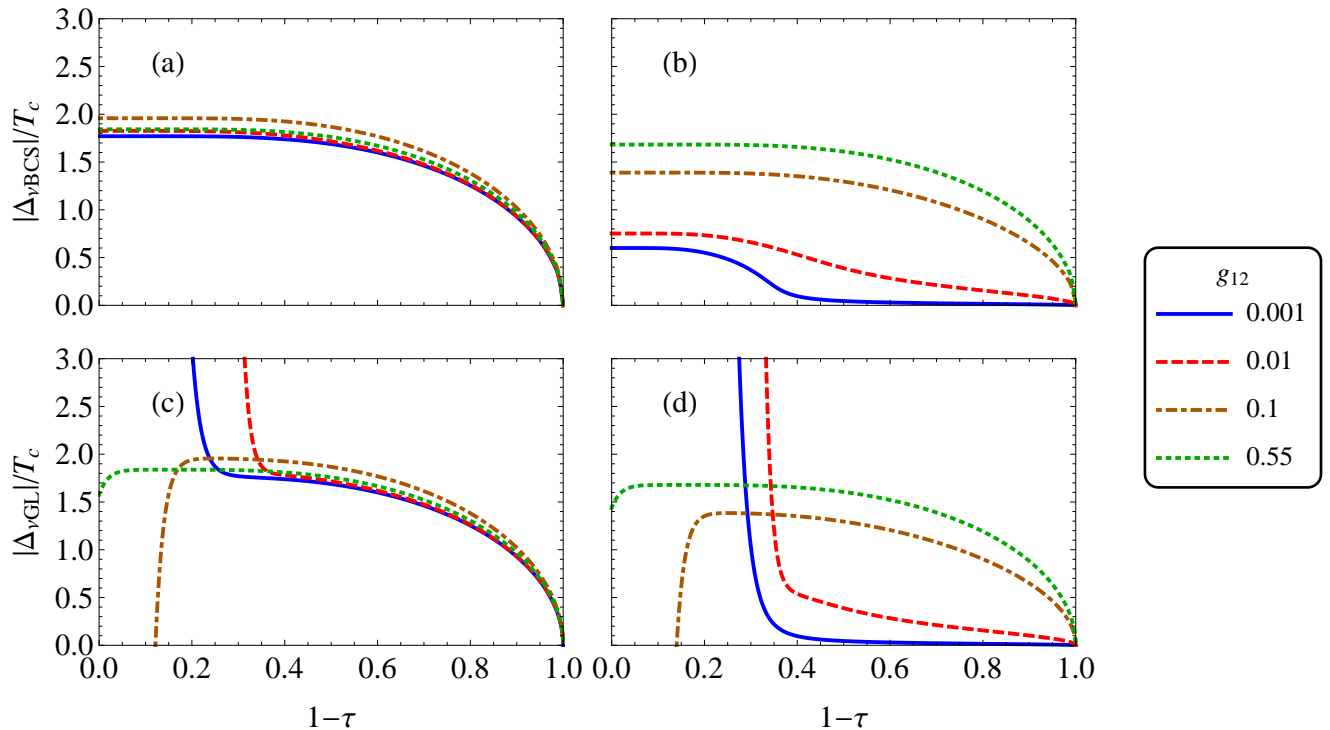}
\caption{(Colour online) Numerical calculations of the BCS gap is compared to the high expansion in the extended GL theory. We use the parameters $g_{11}=0.6$, $g_{22}=0.5$,  $N_{1}(0)=N_{2}(0)=0.3$, $\hbar\omega_{D}=0.09$.
a) BCS solution band 1. b) BCS Solution band 2. c) GL solution band 1. d) GL solution band 2. The GL plots are calculated to order $\tau^{n+1/2}$ where $n=50$.}
\label{fig:BCSGLPlot}
\end{figure*}

\begin{figure*}
\centering
\includegraphics[width=\textwidth]{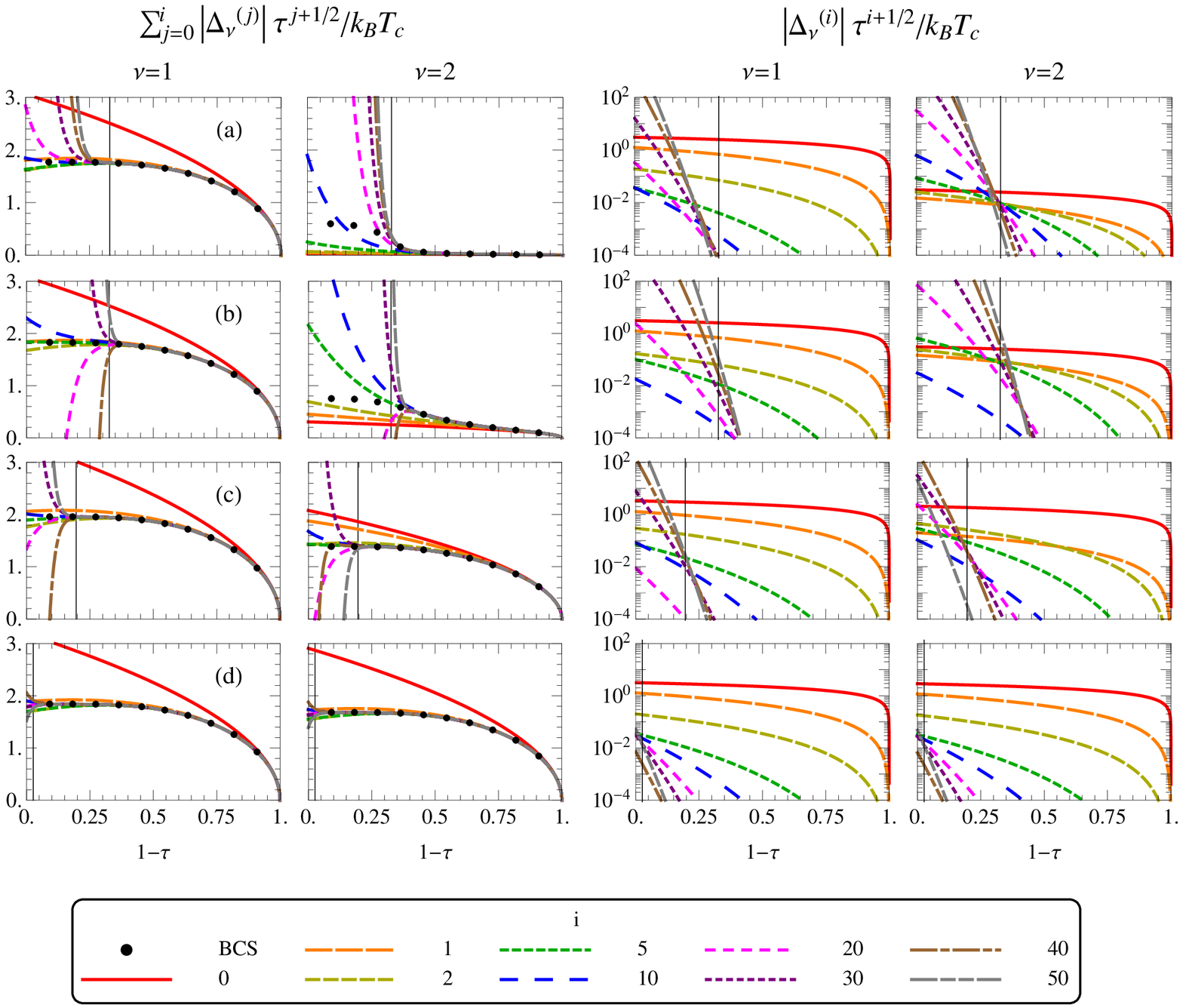}
\caption{(Colour online) The extended GL expansion is compared to a numerical calculation of the full BCS result. 
In all plots we use the parameters $g_{11}=0.6$, $g_{22}=0.5$,  $N_{1}(0)=N_{2}(0)=0.3$, $\hbar\omega_{D}=0.09$.  a) $g_{12}=0.001$, b) $g_{12}=0.01$, c) $g_{12}=0.1$, d) $g_{12}=0.55$. Columns one and three correspond to band 1 while columns two and four correspond to band 2. The first two columns compare a finite sum of terms in the G-L expansion to the full BCS solution. The second two columns show the magnitude of each additional term on a log plot. In all plots the vertical black lines are located at what would be the critical point of the second band in the uncoupled limit, $T_{2c}/T_c$. We see that for $1-\tau \gtrsim T_{2c}/T_c$ the trend is for additional terms to decrease in magnitude, and the series seems to be converging, while for $1-\tau \lesssim T_{2c}/T_c$, the terms tend to grow and the series seems to be diverging. It is seen that the G-L expansion only converges to the BCS result in the region $\tau\lesssim1-T_{2c}/T_{c}<1$. Surprisingly this is true for both the dominant band with small coupling where the dominant band is only mildly perturbed by the interaction, and in the case of intermediate coupling where both bands are convex and there are no sudden increases in the slope of the gap functions.}
\label{fig:ExtendedGLSeries3BigPlot}
\end{figure*}

\begin{figure*}
\centering
\includegraphics[width=\textwidth]{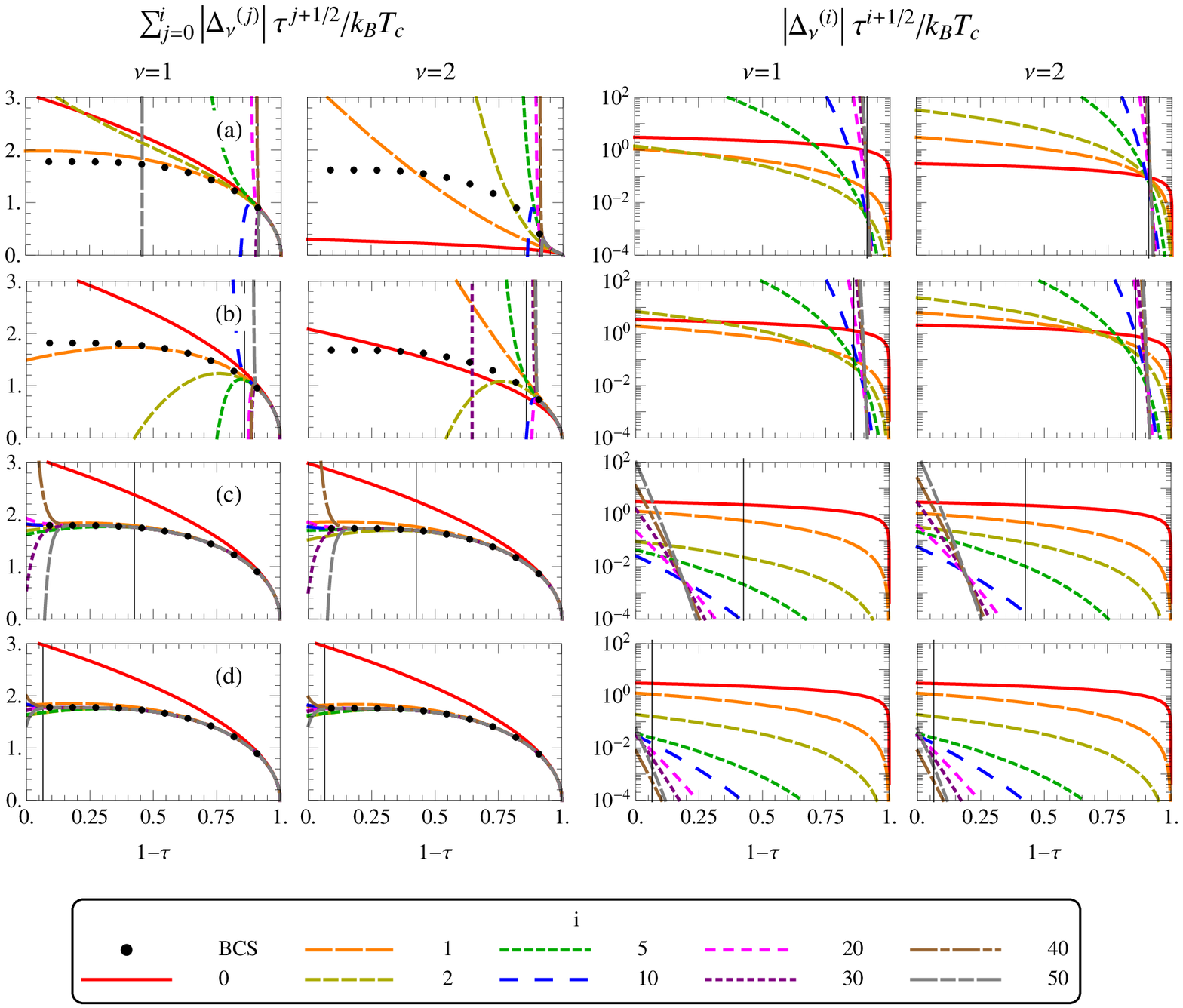}
\caption{(Colour online) The extended GL expansion is compared to a numerical calculation of the full BCS result. 
In all plots we use the parameters $g_{11}=0.6$, $g_{22}=0.59$,  $N_{1}(0)=N_{2}(0)=0.3$, $\hbar\omega_{D}=0.09$.  a) $g_{12}=0.001$, b) $g_{12}=0.01$, c) $g_{12}=0.1$, d) $g_{12}=0.55$. Columns one and three correspond to band 1 while columns two and four correspond to band 2. The first two columns compare a finite sum of terms in the G-L expansion to the full BCS solution. The second two columns show the magnitude of a selection of individual terms on a log plot. In all plots the vertical black lines are located at what would be the critical point of the second band in the uncoupled limit, $T_{2c}/T_c$. Because the critical temperatures are close in the uncoupled limit, the extended GL solution at small coupling only has a very small region of validity. We also see that at larger interband coupling, the location of non-convergent is much lower than the point $T_{2c}/T_c$. Thus, while the non-convergent behaviour is associated with this point at small interband coupling, the location of this point is also a function of $g_{12}$.}
\label{fig:ExtendedGLSeries5BigPlot}
\end{figure*}

In two band GL, things progress in much the same way. However, there is now more a larger range of possibilities due to three parameters in the interband coupling matrix, $g_{\nu\nu'}$, especially the role of the interband interaction, $g_{12}$.

We know from BCS theory that in the limit that the interband coupling goes to zero, the two gaps are independent and each has their own critical temperatures, which we label $T_{1c}$ and $T_{2c}$ respectively.
When the interband coupling is small but nonzero, there is still a large change in the behaviour of the smaller gap near the temperature $T_{2c}$. However the critical temperature of the combined system is an enhancement of the dominant band's critical temperature.

The exact lowest order solution can easily be calculated, with the result
\begin{align}
\Delta_1^{(0)}=&k_BT_c\frac{\sqrt{L_{22}^2N_{1}(0)+L_{12}^2 N_{2}(0)}}{\sqrt{L_{22}^2N_{1}(0)+\frac{L_{12}^4}{L_{22}^2}N_{2}(0)}}\sqrt{\frac{8\pi^2}{7\zeta(3)}}\\
\Delta_2^{(0)}=&k_BT_c\frac{L_{12}\sqrt{L_{22}^2N_{1}(0)+L_{12}^2 N_{2}(0)}}{L_{22}\sqrt{L_{22}^2N_{1}(0)+\frac{L_{12}^4}{L_{22}^2}N_{2}(0)}}\sqrt{\frac{8\pi^2}{7\zeta(3)}}
\end{align}
The higher order terms become increasingly complicated, however the results for specific parameters are calculated numerically to high order.

In figure \ref{fig:BCSGLPlot} we show plots of the BCS solution for a range of values for the interband coupling, $g_{12}$. In a) the first band is plotted, and it is seen that the interband coupling only has a weak effect on the behaviour of this band, while in b), the second gap shows a drastic change as $g_{12}$ increases, especially near $T_{2c}$, the critical temperature of the second band in the noninteracting limit. With the increase of the coupling strength, the large up-swell of the second band near this critical temperature gets washed out, so that at large coupling the plot looks reminiscent of a one band BCS plot.

Plots c) and d) depict the order parameters of band 1 and 2 respectively as calculated using the extended GL formalism derived earlier. For $1-\tau\gtrsim0.3$ the behaviour shown in the GL plots is similar to that of the BCS plots above. However, for $1-\tau\lesssim0.3$ the behaviour of the GL plots is drastically different from the BCS plots, with the difference appearing sooner for smaller $g_{12}$.
The point where the solutions begin to disagree is very close to the location of $T_{2c}$, which in the small coupling limit is $T_{2c}\approx0.33T_c$.
While this finite summation approach does not prove that the series is divergent, it is clear that the sum has not converged in this range for the large number of terms computed. We expect that in general the sum will converge for all $T\gtrsim T_{2c}$, but converge very slowly or diverge for $T\lesssim T_{2c}$. Komendova et al. \cite{Komendova} argue that there is a possibility of hidden criticality near $T_{2c}$ which becomes critical in the limit that the coupling goes to zero. This feature is likely to be associated with the anomalous behavior of the GL gaps near this point, and is expected to prevent the series from converging below this point. 

Surprisingly, while the BCS solution for the first band showed only a weak perturbation with the interband coupling, the non-convergent behaviour seen in the GL solution of the smaller band also affects the dominant band. This occurs for any small non-zero interband coupling, even though the solution converges for all $\tau$ if the interband coupling is zero.

In figure \ref{fig:ExtendedGLSeries3BigPlot} the first two columns show the extended GL of band 1 and band 2 respectively as a function of $1-\tau$ for various $g_{12}$. We can see that as the number of terms included in the expansion is increased, the GL solution departs from the BCS solution, shown as dots, in the region $T\lesssim T_{2c}$, and increasing the number of terms increases this difference. Therefore, with this number of terms, the solution is not converging to the true solution in this range.

The second two columns show the magnitude of each of the terms in the sum on a log plot. In these plots it is shown that there is approximately a pivot point above which the magnitude of the terms decrease, while below this point the magnitude of the terms increase. At the pivot point the magnitude of the terms remains approximately constant. The location of the pivot point is close to the point $T_{2c}/T_c$, especially for small interband coupling. The location of the point $T_{2c}/T_c$ is shown as a vertical black line in the figure.


As $g_{12}$ increases the location of the pivot point seems to move towards $T=0$. However we know that $T_{2c}$ is a constant. A possible reason for this behaviour of the pivot point is that as $g_{12}$ increases, $T_{2c}$ does indeed remain constant, but $T_c$ increases, so that $T_{2c}/T_c$ should move towards $0$ as $g_{12}$ increases. It is this increase in $T_c$ that makes the non-convergent point move towards zero as $g_{12}$ increases.

In figure \ref{fig:ExtendedGLSeries5BigPlot} we have produced a similar plot to figure \ref{fig:ExtendedGLSeries3BigPlot} but where the gaps in the noninteracting limit have similar critical temperatures. With these parameters, the BCS solution shows that the dominant band is almost unperturbed by the interband interaction. At small interband interaction, the second band is weakly perturbed except near $T_c$, however with increasing interband interaction, the second band quickly becomes indistinguishable from the first band. This is expected since the interband interaction causes the two bands to behave as a single band. Since the properties of the two gaps are already similar in the uncoupled limit, only a reasonably small interband interaction is required before the two bands behave like a single band.

When we look at the GL solution, we see that with these parameters and small interband coupling the region of validity of the solution is very tiny.
Even after the inclusion of a very large number of terms, the region where the GL solution has converged to the BCS solution is only in the range $\tau\lesssim0.1$. When the interband coupling is increased, this range of convergence increases significantly. At large coupling, the solution converges over almost the complete temperature range. In this case the two gaps are almost identical.

We see that in the two band case where the two gaps are close to degenerate and the interband coupling is very weak, the GL approximation is only valid in a very small temperature region near $T_c$, and the theory should be applied with care. However, for the case where one band is very dominant, or where the interband coupling is very large, the GL theory performs very well, and converges quickly to the BCS result over a fairly large temperature range.

\section{Conclusion}
In this paper we have reconstructed the relationship of the BCS theory with the GL theory with the limitations developed by Gor'kov in his ground-breaking work. The theory has been restricted to the case of a uniform system, but has been extended to allow multiple bands and large order in $\tau$. This extends on the work of ref \onlinecite{Vagov2} where the authors calculated a similar expansion keeping terms of order $\tau^{3/2}$ in the presence of a magnetic field.


We have shown that in a one band superconductor the $\tau^{3/2}$ correction improves the magnitude of order parameter closer to the BCS value. Higher order corrections for $n\geq1$ in $\tau^{(2n+1)/2}$ improve the agreement with the BCS result except at $T=0$, where the series for the gap appears to be nonconvergent.

In the two band situation, the interband coupling plays a pivotal role in enhancing the smaller order parameter above the $T_{2c}$  value in the BCS model. As the interband coupling increases the point of inflection around $T_{c2}$ heals gradually. At large interband coupling both gaps look similar to a one band solution. The critical temperature of the system evolves smoothly out of the largest critical temperature, $T_{1c}$, and is enhanced by the interband coupling.

In the GL model there are significant differences for both the gaps below $T\lesssim T_{2c}$. The large deviation persists for weaker interband couplings despite including larger $\tau^{(2n+1)/2}$ corrections. This issue is significant when $T_{2c}$ is close to the critical temperature $T_c$. In this case the range of validity of the GL solution can be extremely small. The GL solution to the gaps below $T_{2c}$ is unreliable, and therefore care must be taken when applying the GL model to multiband superconductors. 

When the interband coupling is larger or when one of the gaps is very dominant, the GL solution performs much better and including higher order terms can make the solution close to the BCS value over a large temperature range. Similar to the one band case, the point $T=0$ is nonconvergent in the multiband solution regardless of interband coupling.

In summary we have clearly demonstrated the importance of $\tau^{(2n+1)/2}$ expansion for large $n$ for multiband GL superconductors. This point emphasises the weaker validity of the GL theory for lower temperatures, and especially for applications with small interband coupling. We are of the opinion that any use or misuse of GL theory has to be carefully examined considering its domain of applicability.

\section*{Acknowledgements}
The authors would like to thank the late John Clem, Vladimir Kogan, Alexei Vagov, Francois Peeters, and Nguyen Van Hieu for useful discussions.



\begin{thebibliography}{99}

\bibitem{Yoshida} T. Yoshida, X.J. Zhou, D.H. Lu, S. Komiya, Y. Ando, H. Eisaki, T. Kakeshita, S. Uchida, Z. Hussain, Z-X. Shen, and A. Fujimori, J. Phys:Condens Matter, \textbf{19} (2007) 125209.

\bibitem{Yoshida2} T. Yoshida, M. Hashimoto, I.M. Vishik, Z-X. Shen, and A. Fujimori, J. Phys. Soc. Jpn. \textbf{81}, (2012) 011006.
\bibitem{Knebel} G. Knebel, D. Aoki, J.-P. Brison, L. Howald, G. Lapertot, J. Panarin, S. Raymond, and J. Flouquet,  Physica Staus Solidi  14 June 2013.
\bibitem{Bert} J.A. Bert, B. Kalisky, C. Bell, M. Kim, Y. Hikita, H. Y. Hwang, and K. A. Moler, Nature Physics \textbf{7}, (2011) 767–771.
\bibitem{Gabovich} A. M. Gabovich, A.I. Voitenko, and M. Ausloos, Phys. Rep. \textbf{367}, (2002) 583.
\bibitem{Kimura} N. Kimura and I. Bonalde, Non-centrosymmetric Heavy-Fermion Superconductors, Lecture Notes in Physics Vol. 847 Chapter 2 Ed. E. Bauer  and M. Sigrist (2012).
\bibitem{Stewart} G. R. Stewart, Rev. Mod. Phys.  \textbf{73}, (2001)  797.
\bibitem{Wilson} B. J. Wilson and M. P. Das, J Phys: Condens. Matter \textbf{25} (2013) 425702.
\bibitem{Gorkov} L P Gor'kov, Sov Phys  JETP \textbf{9}  (1959) 1364.
\bibitem{Abrikosov} A. A. Abrikosov, L. P. Gor'kov and I. E. Dzyaloshinski, Methods of Quantum Field Theory in Statistical Physics, Prentice Hall, Englewood Cliffs, NJ (1963), see Chapter 7.
\bibitem{Brandt} E.H. Brandt and M.P. Das J Supercond Nov Magn (2011) \textbf{24} 57
\bibitem{Shanenko} A. A. Shanenko, M. V. Milo\v{s}evi\'{c}, F. M. Peeters and A. V. Vagov, Phys Rev Lett 106 (2011) 047005.
\bibitem{Vagov1} A. V. Vagov, A. A. Shanenko, M. V. Milo\v{s}evi\'{c}, V. M. Axt and F. M. Peeters, PRB \textbf{85}, 014502 (2012)
\bibitem{Vagov2} A. V. Vagov, A. A. Shanenko, M. V. Milo\v{s}evi\'{c}, V. M. Axt and F. M. Peeters, PRB \textbf{86}, 144514 (2012)
\bibitem{Vagov3} N. V. Orlova, A. A. Shanenko, M. V. Milo\v{s}evi\'{c}, F. M. Peeters,  A. V. Vagov and V. M. Axt, , PRB \textbf{87}, 134510 (2013)
\bibitem{Kogan} V. G. Kogan and J Schmalian, PRB \textbf{83} 054515 (2011)
\bibitem{Komendova} L. Komendova, Y. Chen, A.A. Shanenko, M. V. Milo\v{s}evi\'{c}, F.M. Peeters Phys Rev Lett \textbf{108} (2012) 207002. 

\bibitem{Suhl}  H. Suhl, B. T. Matthias and L. R. Walker, Phys. Rev. Lett. \textbf{3}, 552 (1959).
\bibitem{Moskalenko}  V. A. Moskalenko, Fiz. Metal. Metalloved. \textbf{8}, 503 (1959)
\end{thebibliography}
\end{document}